\documentclass[usenatbib]{mnras}

\usepackage{hyperref}
\hypersetup{colorlinks,allcolors=black}
\usepackage{newtxtext,newtxmath}

\usepackage[T1]{fontenc}

\usepackage{csquotes}
\usepackage[dvipsnames]{xcolor}

\DeclareRobustCommand{\VAN}[3]{#2}
\let\VANthebibliography\thebibliography
\def\thebibliography{\DeclareRobustCommand{\VAN}[3]{##3}\VANthebibliography}

\usepackage{graphicx}	
\usepackage{amsmath}	
\usepackage{verbatim}
\usepackage{booktabs} 
\usepackage{tablefootnote} 
\usepackage{threeparttable}
\usepackage[export]{adjustbox}

\title[Optimization and Commissioning of EPIC]{Optimization and Commissioning of the EPIC Commensal Radio Transient Imager for the Long Wavelength Array}

\author[Hariharan et al.]{
Hariharan Krishnan,$^{1}$\thanks{E-mail: hari.krish@asu.edu}
Adam P. Beardsley,$^{2}$
Judd D. Bowman,$^{1}$
Jayce Dowell,$^{3}$ 
Matthew Kolopanis,$^{1}$
\newauthor
Greg Taylor,$^{3}$
and Nithyanandan Thyagarajan$^{4}$ \\
$^{1}$School of Earth and Space Exploration, Arizona State University, Tempe, AZ 85287\\
$^{2}$Department of Physics, Winona State University, Winona, MN 55987\\
$^{3}$Department of Physics and Astronomy, University of New Mexico, 210 Yale Blvd NE, Albuquerque, NM 87106, USA\\
$^{4}$Commonwealth Scientific and Industrial Research Organisation (CSIRO), Space \& Astronomy, P. O. Box 1130, Bentley, WA 6102, Australia
}

\date{Accepted 2023 January 20. Received 2022 December 22; in original form 2022 May 13}

\pubyear{2022}

\begin{document}
\label{firstpage}
\pagerange{\pageref{firstpage}--\pageref{lastpage}}
\maketitle

\begin{abstract}

Next generation aperture arrays are expected to consist of hundreds to thousands of antenna elements with substantial digital signal processing to handle large operating bandwidths of a few tens to hundreds of MHz. Conventionally, FX~correlators are used as the primary signal processing unit of the interferometer. These correlators have computational costs that scale as $\mathcal{O}(N^2)$ for large arrays. An alternative imaging approach is implemented in the E-field Parallel Imaging Correlator (EPIC) that was recently deployed on the Long Wavelength Array station at the Sevilleta National Wildlife Refuge (LWA-SV) in New Mexico. EPIC uses a novel architecture that produces electric field or intensity images of the sky at the angular resolution of the array with full or partial polarization and the full spectral resolution of the channelizer. By eliminating the intermediate cross-correlation data products, the computational costs can be significantly lowered in comparison to a conventional FX~or XF~correlator from $\mathcal{O}(N^2)$ to $\mathcal{O}(N \log N)$ for dense (but otherwise arbitrary) array layouts. EPIC can also lower the output data rates by directly yielding polarimetric image products for science analysis. We have optimized EPIC and have now commissioned it at LWA-SV as a commensal all-sky imaging back-end that can potentially detect and localize sources of impulsive radio emission on millisecond timescales. In this article, we review the architecture of EPIC, describe code optimizations that improve performance, and present initial validations from commissioning observations. Comparisons between EPIC measurements and simultaneous beam-formed observations of bright sources show spectral-temporal structures in good agreement.

\end{abstract}




\section{Introduction}\label{intro}

Modern radio telescopes that rely on synthesis imaging are expanding in size from a few tens of elements in an array to over hundreds to thousands of elements in order to achieve high surface brightness sensitivity, wide field-of-view, wide-operating bandwidth and high-time resolution. This has been facilitated by an exponential growth in digital technology and the availability of commercial off-the-shelf hardware including Field-Programmable Gate Arrays (FPGAs) and general-purpose computing on Graphics Processing Units (GPUs), which have enabled flexibility in the development and the implementation of signal processing firmware \citep{Parsons+2008, Ford2010, Clark2011, Vermij2014, Kocz+2014, Barsdell2014}. Collectively these technologies are transforming radio astronomy into a data-intensive science.

 The technological push is driven by the motivation to understand the physics of a wide-range of astrophysical phenomena. Redshifted 21~cm HI intensity mapping has inspired many new instruments with compact array configurations to achieve high surface brightness sensitivity and wide fields-of-view to make power spectrum measurements of the weak signal on large angular scales. Above $z\approx6$, 21~cm emission and absorption by the intergalactic medium encodes the history of reionization and the properties of the first stars and galaxies.  The first generation of 21~cm arrays for reionization and Cosmic Dawn, including LOFAR, MWA, HERA, PAPER, and OVRO-LWA, use hundreds of compact antenna elements, usually consisting of dipoles \citep{2010AJ....139.1468P, 2018AJ....156...32E} or phased-arrays of dipoles \citep{2013A&A...556A...2V, 2013PASA...30....7T}, but also closely-spaced dishes \citep{HERA2017}.  At lower redshifts, 21~cm emission from galaxies reveals baryon acoustic oscillations in large-scale structure and can be used to study Dark Energy.  The CHIME pathfinder instrument uses cylindrical reflectors with dozens of regularly-spaced dipole feeds per reflector \citep{2014SPIE.9145E..22B}, while plans for next-generation CHORD and HIRAX instruments are based on arrays of hundreds or thousands of small dishes \citep{hirax2016, CHORD2019}.  Similarly, wide fields-of-view coupled with high-time resolution are needed for many transient studies, especially for fast radio bursts \citep{CHIME2018, Cordes+2019, Petroff+2019, Petroff+2021}, meteor radio afterglows \citep{Obenberger+2014, Obenberger+2016}, giant pulse emission from pulsars \citep{Petrova2004, Kuzmin2007}. 
The transient phase space now requires microseconds to millisecond timescale signal processing with high sensitivity instruments in order to detect and characterize sources of interest. The Deep Synoptic Array (DSA) designed for these transient studies has progressed from a 10-element prototype \citep{2019MNRAS.489..919K} to an array of 110 small dishes, with plans to further expand into the thousands.

 The demands of the ever-expanding science and the technological advancements in real-time signal processing have lead to an increase of data volume being generated
which is only set to grow dramatically with the commissioning of new radio telescopes like the Square Kilometer Array (SKA, \citealt{Dewdney+2009}). The conventional techniques of processing the raw signals from antennas using traditional FX/XF~correlators \citep{TMS2017} and the post-processing of the visibility data using specialized offline software tools (e.g. AIPS, \citealt{AIPS1996}; CASA, \citealt{CASA2007}) may no longer be feasible. It is thus necessary to look into novel technological solutions that can process the raw voltage data in real-time, producing science-ready data products, and thereby minimizing the afore-mentioned problems. In view of the above, we have developed a novel correlator design that allows fast imaging with radio arrays consisting of a large number of closely-spaced elements (N $\gtrsim$ 1000). This is referred to as the \enquote{E-Field Parallel Imaging Correlator} (EPIC, \citealt{Thyagarajan+2017, Kent+2019}). 
 EPIC is based on the generalized direct imaging framework called \enquote{Modular Optimal Frequency Fourier} (MOFF, \citealt{Morales2011}). The MOFF formalism takes advantage of the computational speed of the Fast Fourier Transform (FFT).  By gridding sampled antenna voltages onto the aperture plane and applying an FFT to create an image, its computational costs scale primarily as $\mathcal{O}(N_{\rm g} \log N_{\rm g})$, where $N_{\rm g}$ is the size of the aperture plane grid used for the FFT, equivalent to the number of pixels in the resulting image.  In contrast, an FX-correlator calculates the cross-correlation between all antenna pairs before gridding.  This creates a convenient intermediate data product, but its computational costs scale primarly as $\mathcal{O}(N_{\rm a}^2)$, where $N_{\rm a}$ is the number of antennas.  \cite{Thyagarajan+2017} found that the computational costs of EPIC, assumed to be dominated by its spatial Fourier transform, were much lower than for an FX correlator for various planned or possible future telescopes.  EPIC is most suited for densely packed arrays with large numbers of small antennas elements so that $N_{\rm g} \approx N_{\rm a}$, especially for high-time resolution imaging since the total data rate of any imager is inversely proportional to the integration time. 

The Fast Fourier Transform Telescope (FFTT : \citealp{Mao2008, Tegmark2009, Tegmark2010}) or the Omniscope, is a closely-related direct imaging approach  that has been tested on the Basic Element for SKA Training II array (BEST : \citealp{Foster2014}) and MITEoR experiment \citep{Zheng2014}. The FFTT assumes a redundant configuration for the array layout and requires all antenna elements to be identical.  EPIC does not apply these constraints and can be used on heterogeneous arrays with non-identical antennas.  The primary advantage is that EPIC can be applied to arbitrary array layouts and it can include individual antenna beam patterns in its gridding kernels, making it more generic and better suited for high-dynamic range imaging.  EPIC and FFTT have the same scaling with $N_{\rm g}$, but EPIC incurs additional computational overhead in the gridding and may not be able to apply some FFT optimizations for specific hierarchical array layouts.

EPIC is currently being developed and tested as a commensal imaging back-end for the Long Wavelength Array (LWA : \citealp{Ellingson+2009, Henning+2010}). LWA is an excellent testbed for EPIC as it consists of hundreds of antennas in a very compact pseudo-random configuration that yields high-dynamic range imaging.  The array also has well-modeled beam patterns and good phase stability across its operating frequency range, reducing the need for time-varying calibration solutions. As with other direct imagers, calibration remains an open area of development for EPIC.  Unlike FX-correlators where calibration information can be applied to the visibilities offline before the final imaging stage, direct imaging approaches require most calibration information to be applied in real-time.  A feedback calibration scheme called EPICal was proposed and demonstrated for EPIC through software implementation using archived data by \cite{Beardsley+2017}. More work is needed to implement EPICal into the operational system. We note that a similar holographic calibration scheme \citep{Kiefner2021} and reduced-redundant baseline calibration schemes \citep{Gorthi2021} were recently proposed for phased-array telescopes that may apply to FFT correlators like EPIC.

 The first GPU-based implementation and deployment of EPIC was carried out by \cite{Kent+2019}. The bandwidth processed per GPU in real-time was limited. This was attributed to the memory resource capacity of the hardware and its usage efficiency by the firmware associated with some key components of EPIC. In this article we discuss the hardware upgrade and the low-level optimization of two critical components of EPIC and the performance improvements achieved with it through commissioning observations. We begin with a brief review to the MOFF formalism and a description of the GPU implementation of the EPIC architecture in Section~2. In Section~3, the software optimizations performed on the critical aspects of the pipeline are detailed and demonstrated through performance measurements. In Section~4, we showcase observational results from EPIC that are validated through simultaneous beam-formed observations. Finally, we summarize and conclude by discussing the future prospects for EPIC in Section~5.

\section{E-Field Parallel Imaging Correlator}

 Here we provide a sketch of the mathematical framework behind the MOFF algorithm
as implemented by EPIC. Detailed derivations can be found in \citet{Morales2011} and \citet{Thyagarajan+2017}. In interferometric arrays the fundamental measurement of an FX correlator is 
the cross-correlation of the voltages between antenna pairs, recorded as a measure 
of the spatial coherence function called the \emph{visibility}. Restricting ourselves to
a planar array, the visibility is related to the intensity distribution in the sky 
given by the Fourier relationship \citep{TMS2017, Kent+2019},
\begin{multline} \label{eq:visibility}
 V_{ab} = \left<E_a(t)E^*_b(t)\right>_t \\ = \int A(l,m)I(l,m) \exp\left[-2\pi i(ul + vm)\right]\, d\Omega \end{multline}
Here, $(l, m)$ are the direction cosines in specific sky positions, $A(l,m)$ is the beam response of the antenna, $I(l,m)$ is the intensity from the sky at the specific sky positions, $E_a(t)$ and $E_b(t)$ are the measurements from 
antennas $a$ and $b$ at time $t$, $V_{ab}$ represents the visibility matrix and $(u,v)$ 
is the baseline co-ordinate between antennas in the interferometer.

Forming a sky image from the visibility amounts to inverting this Fourier relationship. This is often done by gridding the visibilities to regularly spaced grid cells to leverage the efficiency of an FFT. The estimate of the sky intensity can be expressed as
\begin{equation} \label{eq:fx_image}
I_D(l,m) = \mathcal{FT}^{~-1}\left[(B_{ab} * V_{ab})(u, v)\right],
\end{equation}
where $B_{ab}(u, v)$ is a baseline-dependent gridding kernel (in the optimal mapmaking framework this is taken to be the transpose conjugate of the Fourier transform of the beam response pattern), and $I_D(l,m)$ is called the \emph{dirty image} as a deconvolution has not been performed.

The MOFF algorithm is formulated by recasting the gridded visibilities as a
convolution of gridded antenna measurements.
\begin{equation}
\left(B * V_{ab}\right)(u, v) = \left<\left[W_a * E_a\right] * \left[W_b * E_b\right]^*\right>_t(u, v)
\end{equation}
Here $W_{a}$ is the gridding kernel for antenna $a$. We then use the 
multiplication-convolution theorem to Fourier transform the gridded antenna 
measurements, and multiply in the Fourier (image) domain.

\begin{equation} \label{eq:moff_image}
 I_D(l,m) = \left< \left|\mathcal{FT}^{-1}\left[ \left(W_a * E_a\right)  \right] \right|^2 \right>_t
\end{equation}

\noindent

It follows from equation \ref{eq:moff_image} that the sky representation can be 
reconstructed by directly gridding the measured electric field patterns of the 
individual antennas onto a regularly gridded aperture-plane followed by a spatial 
Fourier transform. This step essentially eliminates the cross-correlation 
operation which forms the crux of the FX-correlator that measures visibilities. 
The gridded complex voltage patterns are then cross-multiplied between polarizations 
in the image plane and accumulated over a period of time to produce dirty images.

\subsection{Deployment of EPIC at LWA-SV}

The LWA is a low-frequency radio interferometer observing over the frequency range of 10$-$88 MHz, with operational stations located at the Karl G. Jansky Very Large Array site \citep[ LWA1]{Taylor+2012} and at the Sevilleta National Wildlife Refuge \citep[ LWA-SV]{Cranmer+2017}, both in New Mexico \citep{Ellingson+2009, Henning+2010}. Each station of the LWA consists of 256 dual-polarized dipole antennas in a pseudo-random arrangement within a compact elliptical aperture measuring 110~x~100~m. 

Considering the compact configuration and the phase stability of the LWA station, EPIC has been deployed and tested on the LWA-SV station \citep{Kent+2019}. The low-pass filtered and amplified analog signal from each individual antenna in the array is transmitted through co-axial cable to an electronics shelter. Within the shelter the signal is further filtered and digitized using the CASPER ADC16x156-8 digitizer that is attached to a ROACH2 board. Onboard the ROACH2, the digitized time series is then channelized to 4096 channels at a frequency resolution of 25~kHz and time resolution of 40~$\mu$s. The data is further requantized to 4+4 complex integers that is packetized and routed over a 10/40~GbE network to a cluster of seven general purpose machines. Each machine is equipped with two Intel Xeon E5-2640 v3 CPUs with 8 cores each\footnote{\url{htps://www.intel.com/content/www/us/en/products/sku/83359/intel-xeon-processor-e52640-v3-20m-cache-2-60-ghz/specifications.html}}, a 40~GbE network interface card and two NVIDIA GTX 980 (Maxwell) GPUs.

\subsection{EPIC Architecture}\label{epicarch}

The EPIC architecture has been implemented using the high performance streaming framework, Bifrost \citep{Cranmer+2017}, that consists of independent modules implemented as low-level C++ libraries with a C API which is the \enquote{back-end} and a Python-based high-level interface with direct wrappers to the C++ libraries which is the \enquote{front-end}. A pipeline in Bifrost is based on the concept of functional blocks that perform specific operations on the data with its front-end and back-end components supporting high-speed parallel computing on GPUs enabled by NVIDIA's Compute Unified Device Architecture (CUDA) libraries. Bifrost also supports the generation and implementation of GPU codes at run time using NVRTC which is a runtime compilation library for CUDA C++. A  generic mapping function, called the  the \enquote{map} block, is included in the Bifrost front-end that takes a string of CUDA C++ source code to create a new GPU kernel during run time. Each of these Bifrost blocks are connected through high-speed memory ring buffers that enable the streaming of data across blocks. The data are ingested into the blocks that processes them and load output buffers for subsequent blocks in the pipeline, until the input buffer is emptied or the pipeline is shut down. Many of the standard signal processing such as digital filters, Fourier transforms, etc. are already implemented in Bifrost.

\begin{figure}
	\includegraphics[height=0.6\textwidth, width=0.75\columnwidth, center]{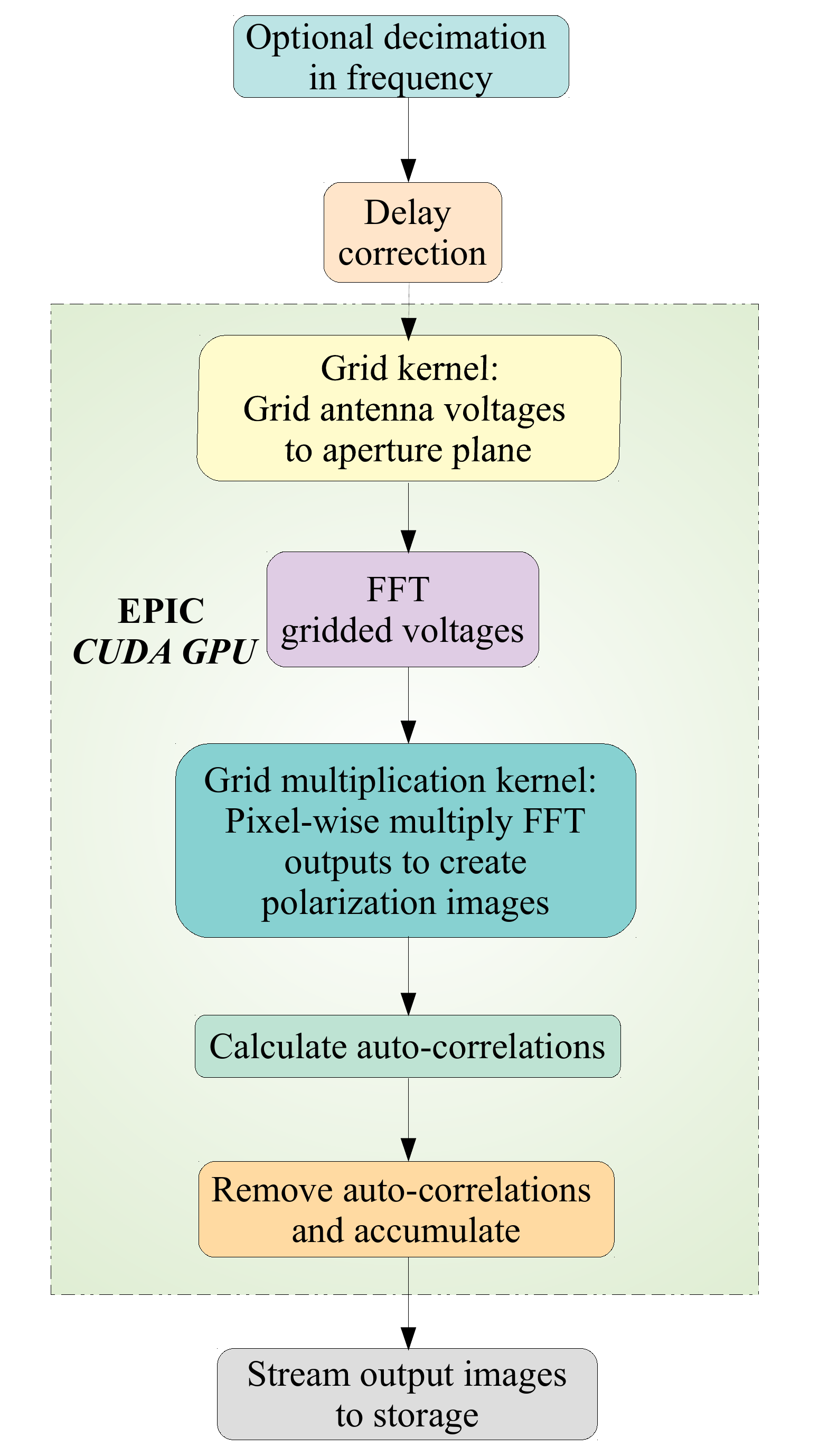}
    \caption{Block Diagram representing the signal flow within EPIC}
    \label{fig:epicblock}
\end{figure}

As illustrated in the block diagram in Figure \ref{fig:epicblock}, EPIC consists of Bifrost blocks that run on both the host CPU and the GPU. On the host CPU, the decimation block slices the incoming frequency domain data to a specified number of frequency channels to be processed by the subsequent blocks in EPIC. Following the frequency slicing, the delay correction block compensates the geometrical delay across all antennas in the interferometer for every time step with respect to the reference antenna or phase center of the array across all frequency channels. This data is then ingested to the global memory, i.e. the Random Access Memory (RAM) of the GPU, which is further processed. Multiple instances of the Bifrost blocks on GPU are simultaneously initiated to achieve parallel processing of the data across time and frequency. The delay corrected voltage signals are then mapped to a regular grid on the aperture plane for every time step by the gridding block to generate two-dimensional voltage grids for the two orthogonal polarizations. These grids are then spatially Fourier transformed in the FFT block to generate complex voltage grids in the image plane. These are then multiplied pixel-wise in a grid multiplication block generating polarization image grids. The map block in Bifrost is then used to implement a kernel that generates auto-correlation grids from individual antenna voltages. The image grids are further accumulated for a specified integration time following the subtraction of the auto-correlation grid. Finally, the save block runs on the host server streaming the image products retrieved from the accumulation block onto the disk for storage. Thus, EPIC essentially synthesizes an aperture on-the-fly eliminating the intermediate cross-correlation step normally involved in conventional FX~correlators and thereby reducing the computational scaling from $\mathcal{O}(N_{\rm a}^{2})$ to $\mathcal{O}(N_{\rm g}\log_{2}$ $N_{\rm g})$.

\subsubsection{Hardware Upgrade}

\begin{figure}
	\includegraphics[width=\columnwidth, center]{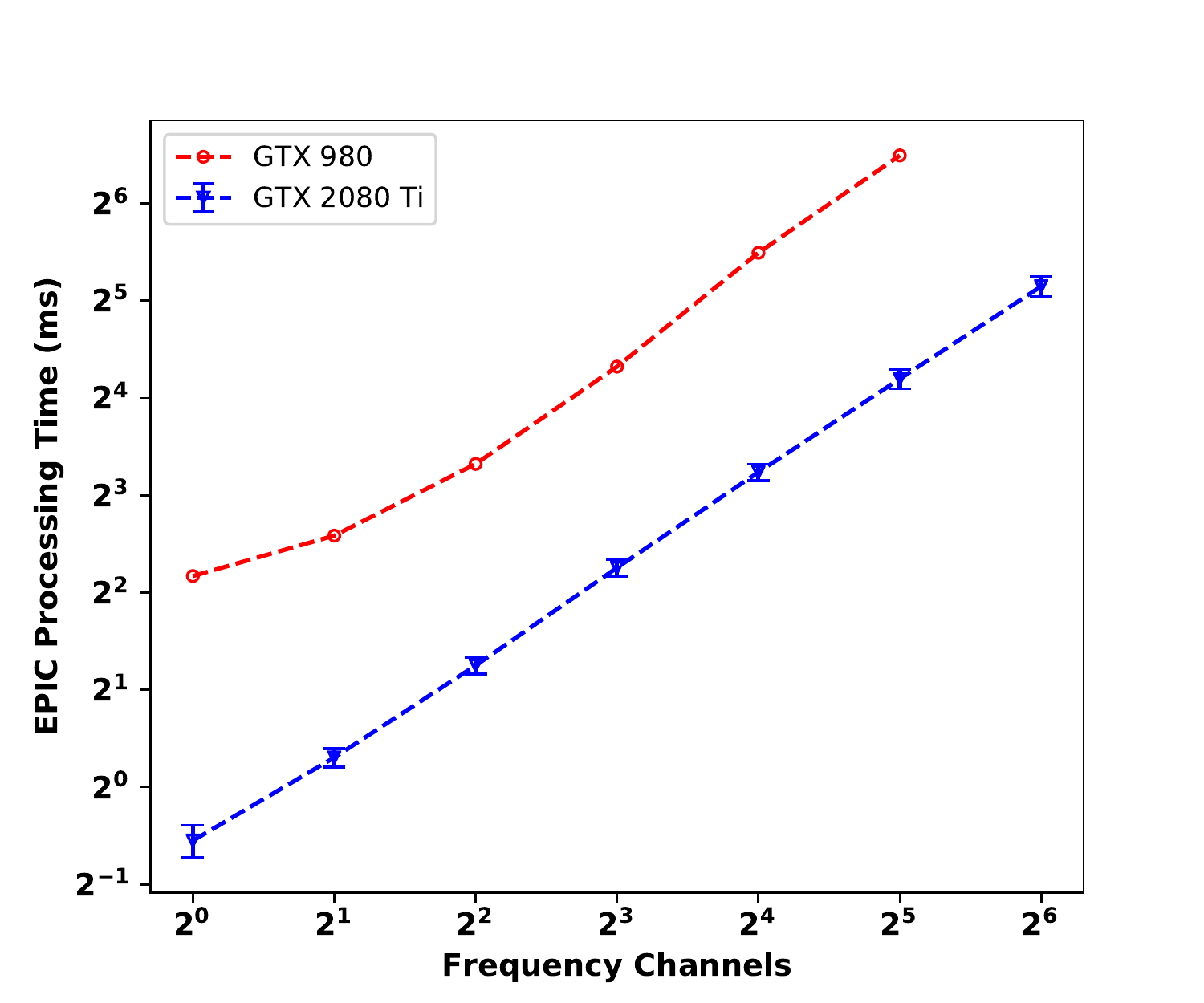}
    \caption{Comparison of the processing time of EPIC on the two GPUs, NVIDIA GeForce GTX 980 \citep{Kent+2019} and RTX 2080 Ti, for increasing number of frequency channels with constant grid size of 32$\times$32 and time gulp of $\sim$50 ms for full polarization. Note that here we use the original code for EPIC from the first deployment.}
    \label{fig:gpuvsgpu}
\end{figure}

  The instantaneous bandwidth achieved with the initial deployment of EPIC on LWA-SV was limited to $\sim$100 kHz per GPU. In the concluding remarks, \cite{Kent+2019} noted that hardware and software upgrades were necessary to increase performance in terms of the bandwidth processed per GPU. In view of the above, a dedicated commensal server machine with new GPU hardware has since been installed at the LWA-SV station. The current system now consists of a single server with two Intel Xeon Silver 4210\footnote{\url{https://www.intel.com/content/www/us/en/products/sku/193384/intel-xeon-silver-4210-processor-13-75m-cache-2-20-ghz/specifications.html}} processors with 10 cores each, 96 GB of RAM, two NVIDIA GeForce RTX 2080 Ti GPUs, and a 40 GbE network interface. The computer is connected to the data switch in the Advanced Digital Processor (ADP) through a dedicated 40 GbE optical link \citep{lwamemo214}. A comparison of the two GPU cards in Table \ref{Tab:gputab} shows that the new 2080 Ti's have twice the memory bandwidth and compute capacity in comparison to the earlier NVIDIA GTX 9080s. A comparison of the total processing time associated with the GPU modules of EPIC profiled on the two GPUs shown Figure \ref{fig:gpuvsgpu} indicates speedup by a factor of four.
 
\begin{table}
\begin{threeparttable}
\normalsize
	\centering
	\label{Tab:gputab}
	\begin{tabular}{lcc}
		\hline
		 Parameters & GTX 980 & RTX 2080Ti \\
		\hline
		Number of Cores & 2048 & 4352 \\
		GPU Clock (MHz) & 1127 & 1350 \\
		Streaming Multiprocessors & 16 & 68 \\
		Memory Bandwidth (GB/s) & 224.4 & 616 \\
		FP32 Performance (TFLOPS) & 4.891 & 13.45 \\
		FP64 Performance (GFLOPS) & 155.6 & 420.2 \\
		\hline
	\end{tabular}
 \caption{Comparison of performance parameters between the GPU cards NVIDIA GeForce GTX 980\tnote{1} and NVIDIA GeForce RTX 2080 Ti\tnote{2}}
	
\begin{tablenotes}
\item[1] \url{https://www.nvidia.com/en-us/geforce/gaming-laptops/geforce-gtx-980/specifications/}
\item[2] \url{https://www.nvidia.com/en-us/geforce/gaming-laptops/geforce-gtx-980/specifications/}
\end{tablenotes}
\end{threeparttable}
\end{table}

\section{Optimization of EPIC}

  \cite{Kent+2019} noted the unique computational challenges of EPIC could further benefit through software optimizations of its GPU kernels. In this regard, we have implemented code modifications that address GPU memory management and data transfer issues within some of the critical blocks of EPIC to improve the overall efficiency and performance in terms of real-time operating bandwidth processed per GPU. In this section we summarize the motivation for two specific optimizations and the changes that have been implemented.
 
\subsection{Gridding Kernel}

  Gridding the antenna voltages is one of the crucial steps involved in EPIC. A regular $(x,y)$ grid with a spacing $\leq\lambda/2$ is generated in a coordinate system with the known antenna locations. An efficient k-dimensional tree algorithm is then used to create a nearest-neighbour mapping \citep{ManeeMount1999} of the antenna footprints to the grid locations. A gridding kernel is implemented as a custom Bifrost block using the efficient work distribution strategy for high-speed convolution and gridding described by \citet{Romein2012}. The gridding kernel could, in general, incorporate multiple effects as a single convolution kernel, like the illumination pattern (A-projection, \citealt{Bhatnagar+2008}) and the wide-field correction ($w$-projection, \citealt{Cornwell+2005}). The \enquote{Romein Gridder} essentially reduces the overall GPU memory bandwidth utilization through explicit memory store operations and performs the convolutional mapping of electric fields onto the grid. The frequency domain voltage series from individual antennas that have been corrected for band-pass delay are convolved with the gridding kernel and gridded on the regular two-dimensional ($x,y$) grid pattern in the aperture plane.
 
  In the initial deployment \cite{Kent+2019} had implemented the Romein gridder achieving functional parallelism for the gridding operation on the GPU. However, the memory and resource utilization of the Romein gridder were sub-optimal and well below the compute capacity of the GPU which degraded performance of the gridding. In \citet{Hariharan+2020}, we showed that the performance efficiency of the Romein griddder could be improved by optimizing memory access \citep{CUDA2012}. Thus the Romein gridder was modified and a new gridding kernel {\tt VGrid} is now implemented as a Bifrost block that is specifically optimized and used for EPIC. 
 
\begin{figure}
	\centering
	\includegraphics[width=\columnwidth, center]{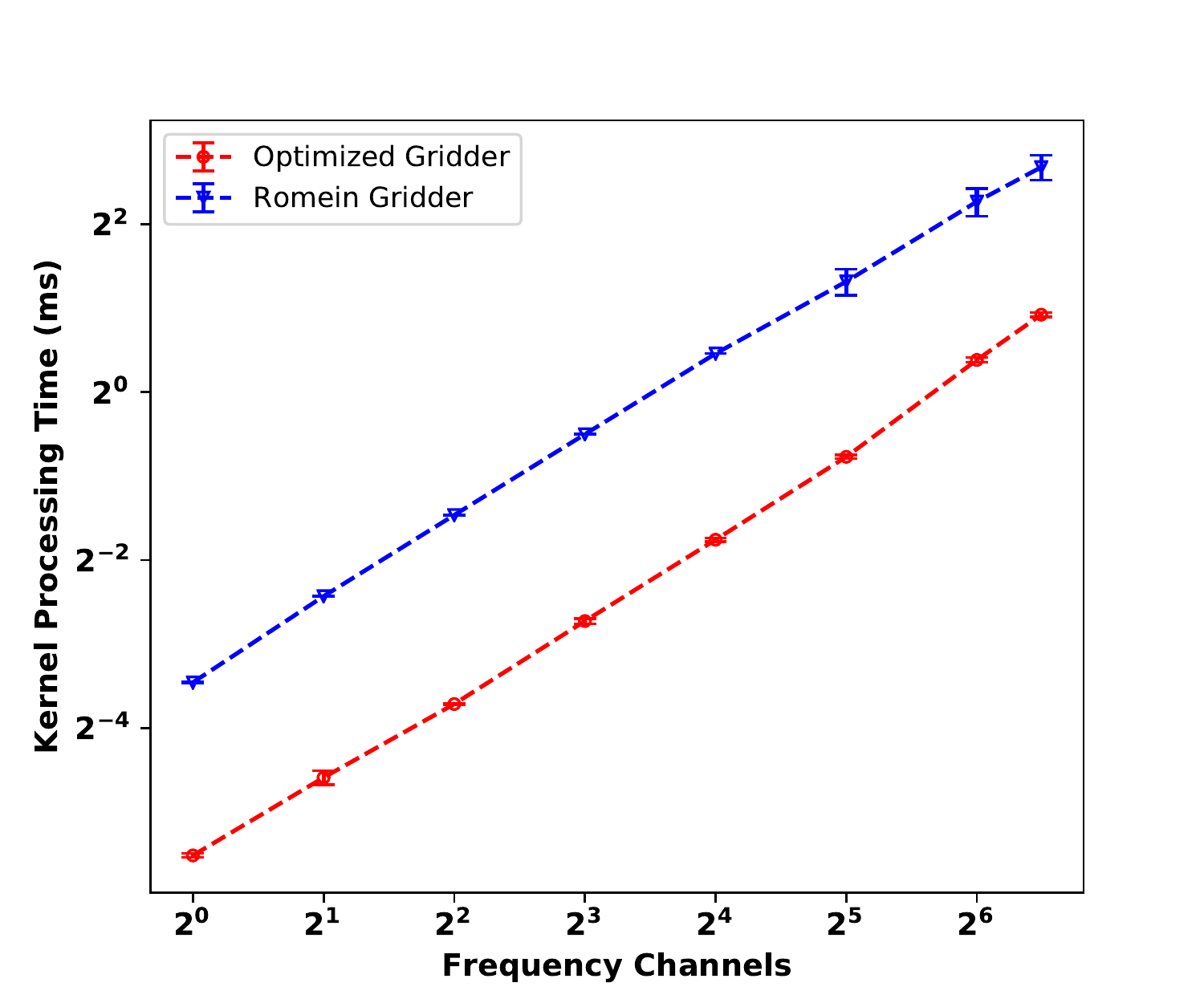}
   
	\includegraphics[width=\columnwidth, center]{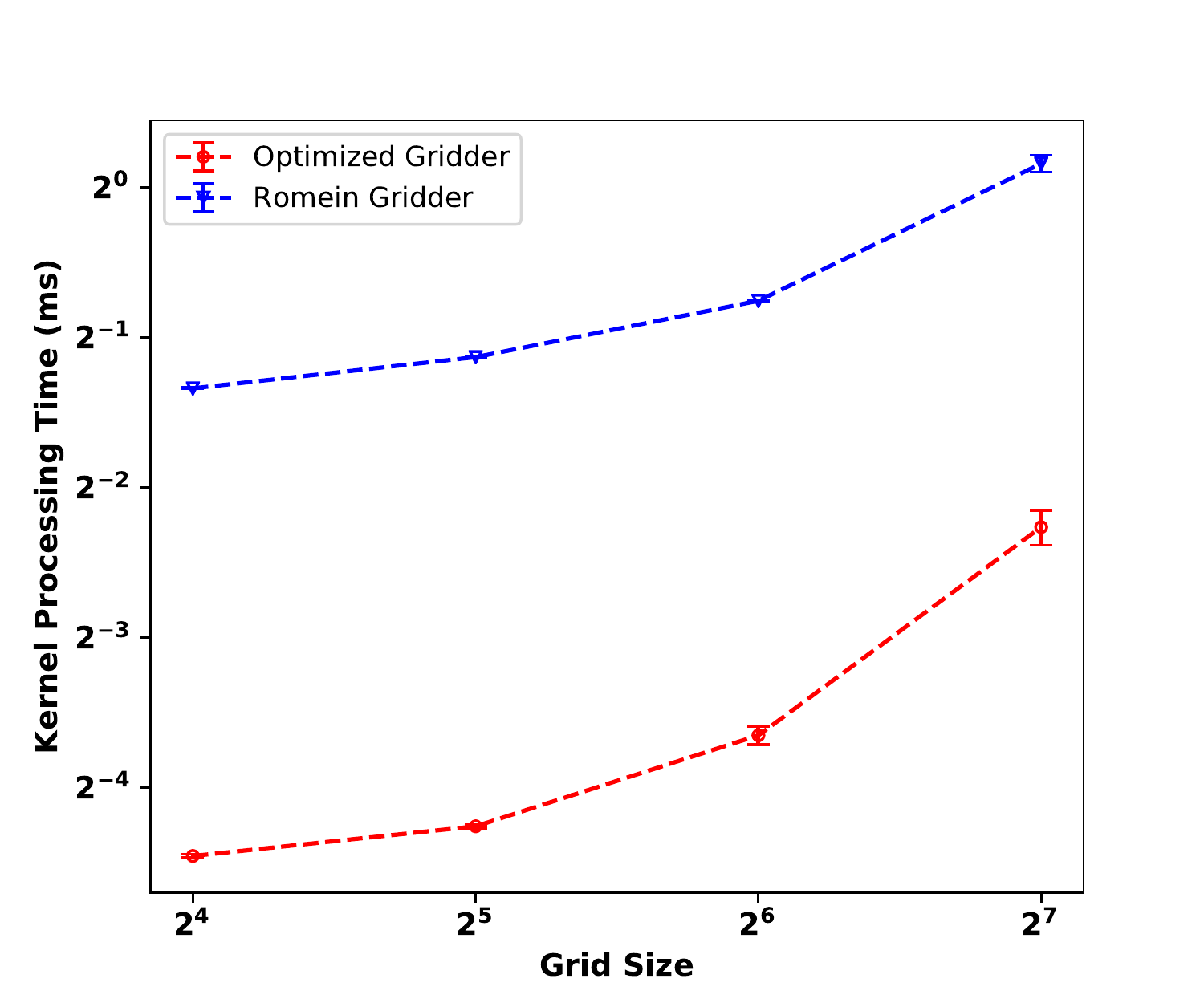}
    \caption{Comparison of the Romein method of gridding and the optimized gridding kernel performance for a time gulp of $\sim$40~ms. The top panel plots the processing time against increasing number of frequency channels with a constant grid size of 64~x~64 (labeled as $2^6$ in the bottom panel).  The bottom panel plots the processing time against increasing grid-size when processing four ($2^2$) frequency channels.}
    
    \label{fig:vgrid}
\end{figure}

  Both the Romein and {\tt VGrid} voltage gridding kernels were simultaneously profiled on the NVIDIA GTX 2080 Ti. The profiling is performed by varying the grid size and the number of frequency channels used in the computation up to the limit imposed by the memory capacity of the GPU.  The number of frequency channels is tunable by a decimation block implemented before the main EPIC processing flow, as described in Section~\ref{epicarch} and shown in Figure~\ref{fig:epicblock}. Figure~\ref{fig:vgrid} shows the comparison of the processing time for different grid sizes and number of frequency channels. In the tests, the time gulp is fixed at $\sim$40~ms. The time gulp corresponds to the number of F-engine iterations that are processed by EPIC in a single cycle of the data flow loop shown in Figure~\ref{fig:epicblock}. The improved performance of the optimized gridding kernel is about a factor of four compared to the original Romein implementation.

\subsection{Grid-Multiplication Kernel}

  The grid-multiplication kernel in EPIC performs cross-multiplication of the complex voltage grids to produce four image grids representing the polarization products. It was implemented by \cite{Kent+2019} using the convenient map block in Bifrost (described in Section~\ref{epicarch}).  However, this suffered similar memory utilization issues as the Romein gridder and could not be tuned at the low-level for optimization due to its very generic form in the map block. To optimize this calculation, a new kernel called {\tt XGrid} is now included as a Bifrost block in EPIC that replaces the map block implementation of the grid-multiplication function. The {\tt XGrid} block performs pixel-wise multiplication of the complex voltage grids V$_X$ and V$_Y$ for the two orthogonal polarizations $X$ and $Y$ generating the full (or partial) polarimetric (V$_X$V$_X$$^*$, V$_Y$V$_Y$$^*$, V$_X$V$_Y$$^*$, and V$_Y$V$_Y$$^*$) image products. The cross-multiplication of the grids is implemented as a pixel-wise Hadamard product mathematically represented in equation \ref{eq:5} showing the element-wise multiplication of two matrices $A$ and $B$ of $m \times n$ dimensions.
 
  \begin{equation} \label{eq:5}
A \circ B = (a_{ij}\cdot b_{ij}) = 
\begin{pmatrix} 
a_{11} \cdot b_{11} & \cdots & a_{1n} \cdot b_{1n} \\
\vdots & \ddots & \vdots \\ 
a_{m1} \cdot b_{m1} & \cdots & a_{mn} \cdot b_{mn} 
\end{pmatrix}
\end{equation}

\begin{figure}
    \centering
 	\includegraphics[width=\columnwidth, center]{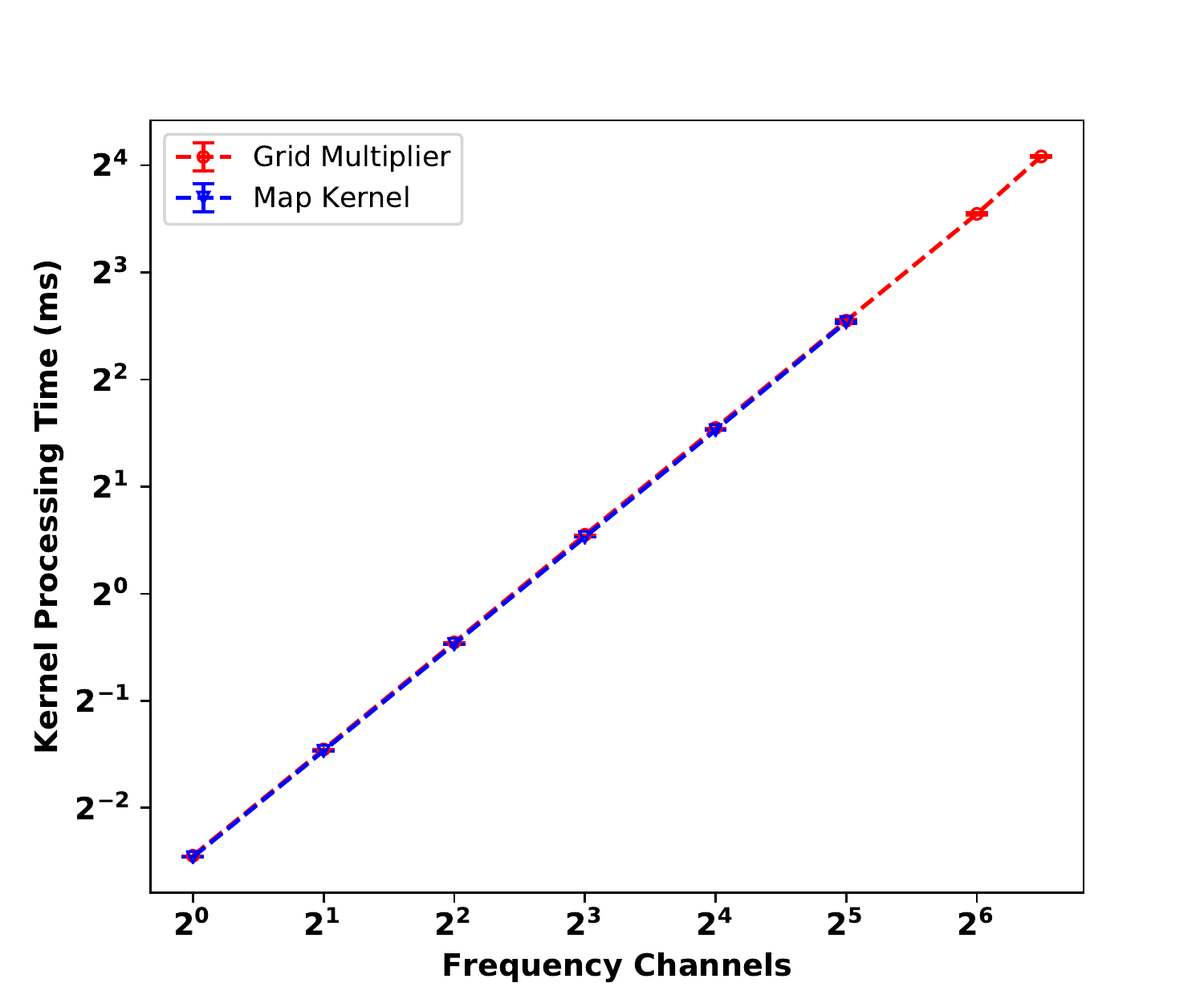}
	\includegraphics[width=\columnwidth, center]{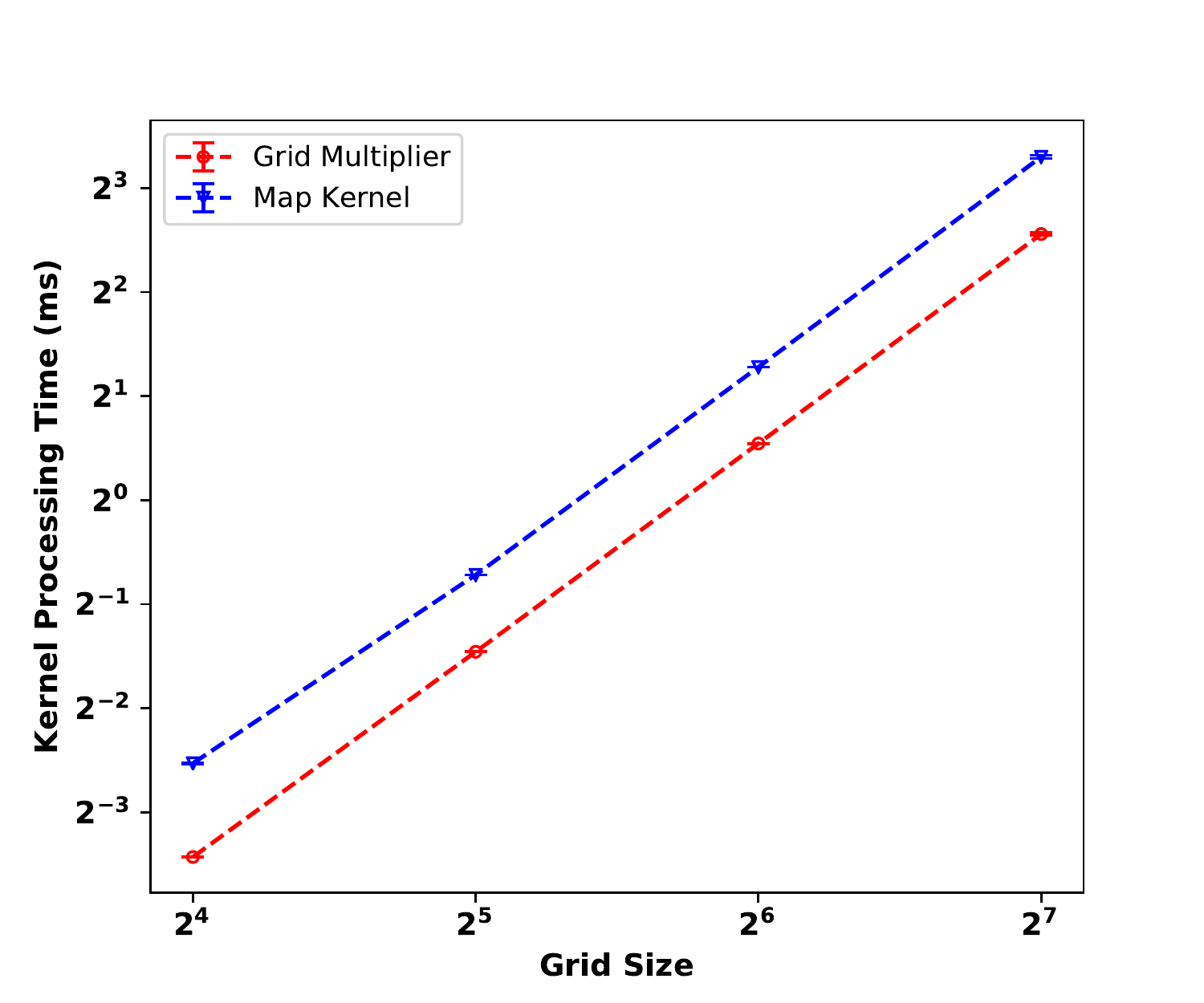}
 	 \caption{Comparison of the grid multiplication implemented using the Bifrost map block and the optimized Grid multiplier for a time gulp of $\sim$40~ms.  The top panel plots the processing time with respect to the number of frequency channels with a constant grid size of 64~x~64 (labeled as $2^6$ in the bottom panel) for single polarization. The optimized kernel makes better use of memory and is able to process more frequency channels, extending the curve in the upper right. The bottom panel plots the processing time against increasing grid size when processing four ($2^2$) frequency channels for full polarization. In the full polarization case, the improved memory utilization leads to increased parallelism across polarization, improving the performance speed of the optimized Grid multiplier in comparison to the map block.}

      \label{fig:xgrid}
 \end{figure}

The top panel of Figure \ref{fig:xgrid} shows a comparison of processing time for the two blocks for varying frequency channels with a constant grid-size of 64~x~64 (2$^6$) and time gulp of $\sim$ 40 ms in single polarization mode. It can be seen that though the kernel performances remain the same we are able to process up to four times as many frequency channels for similar grid- and gulp- sizes which is likely due to optimization of the local memory usage within the GPU. The bottom panel of Figure \ref{fig:xgrid} shows a similar comparison for various grid-sizes with the same time gulp and four (2$^2$) frequency channels, but in full-polarization mode, demonstrating a factor of two improvement in speed.

With the above hardware upgrade and firmware optimizations to critical blocks of gridding and grid-multiplier, EPIC is now significantly improved from the first deployment as indicated in Table \ref{Tab:processtab} which shows an approximate breakdown of processing time of individual blocks in EPIC.

\begin{table}
\begin{threeparttable}
\normalsize
	\centering
	
	\label{Tab:processtab}

	\begin{tabular}{lcc}
		Blocks & \multicolumn{2}{c}{Processing Time} \\
		~ & \multicolumn{2}{c}{[\% of gulp time]} \\
		~ & This Work & \citep{Kent+2019}\\ 
		      \hline 
		      \hline
	    Decimation & 2 & 2 \\
	    Data Transport \tnote{1} & 4 & 4 \\
	    EPIC-GPU \tnote{2} & 19 & 90 \\
		Image Storage & 1 & 4  \\
		\hline
	\end{tabular}
    \caption{Comparison of representative approximate breakdown of processing time of each of the EPIC block as a fraction of time gulp for a grid size of 64~x~64 (2$^6$), 2048 40$\mu$s time samples and 8 frequency channels with \protect\cite{Kent+2019}}
	\begin{tablenotes}
	\item[1]{Data transport is the total memory transfer time between the GPU and the host server}
	\item[2]{Here EPIC-GPU includes gridding, FFT, grid-multiplication, auto-correlation removal, accumulation}
	\end{tablenotes}
	\end{threeparttable}
\end{table}

\section{Observations}

 Commissioning observations with EPIC were performed over several months beginning in August 2021. Through a number of long duration trial observing runs we empirically arrived at optimal values for the time gulp, grid size, and number of channels that can be supported by the compute capacity and memory bandwidth of the GPU to ensure seamless image acquisition without data packet loss. The optimal values of $\sim$40~ms, 64~x~64, and 90, respectively, yield a processed bandwidth of $\sim$1.8~MHz per GPU.  Below we provide representative examples of EPIC observations along with comparisons to simultaneous beam-formed observations created by the standard LWA-SV processing pipeline running commensally. 

Antenna-based gain corrections are presently not implemented in EPIC and are still under development as mentioned in Section~\ref{intro}. Here we perform flux calibration of the images by normalizing the image matrix by the brightest pixel corresponding to a bright ``A-team'' source in the field. The flux densities are then estimated for the source using models from \citet{PB2017} and scaled by the directive gain of the LWA dipoles \citep{lwamemo178,lwamemo202}. 

\subsection{Solar Radio Bursts}

\begin{figure}
    \centering
	\includegraphics[width=\columnwidth]{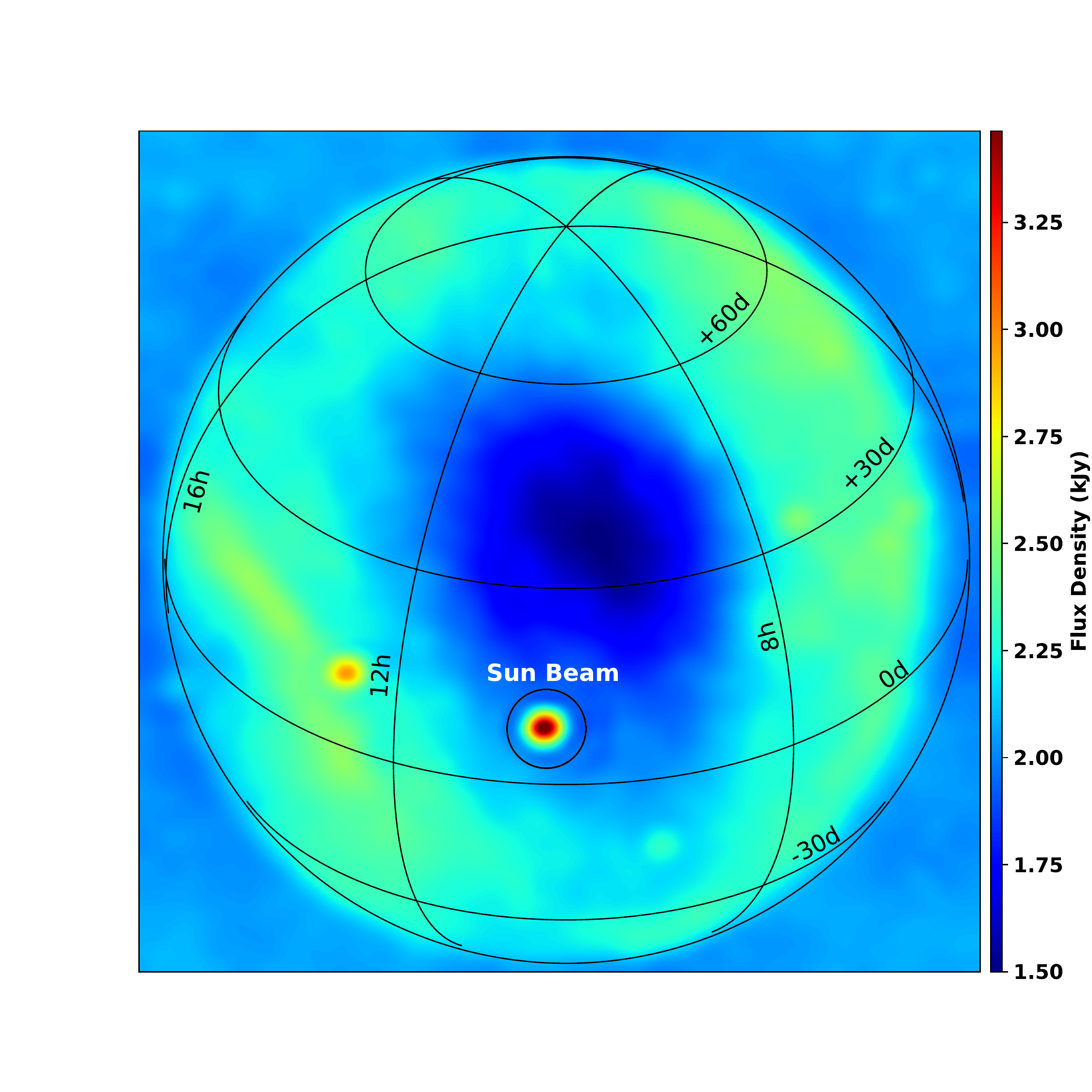}
    \caption{Snapshot full-sky image using EPIC at a time integration of $\sim$1 second over a band of 1.8~MHz center at 43~MHz. The black circle marks the location of the simultaneous beam-formed observations of the Sun acquired by the standard LWA-SV pipeline for dynamic spectroscopy.} 
    \label{fig:epicsun}
\end{figure}

\begin{figure}
    \centering
	\includegraphics[width=\columnwidth, center]{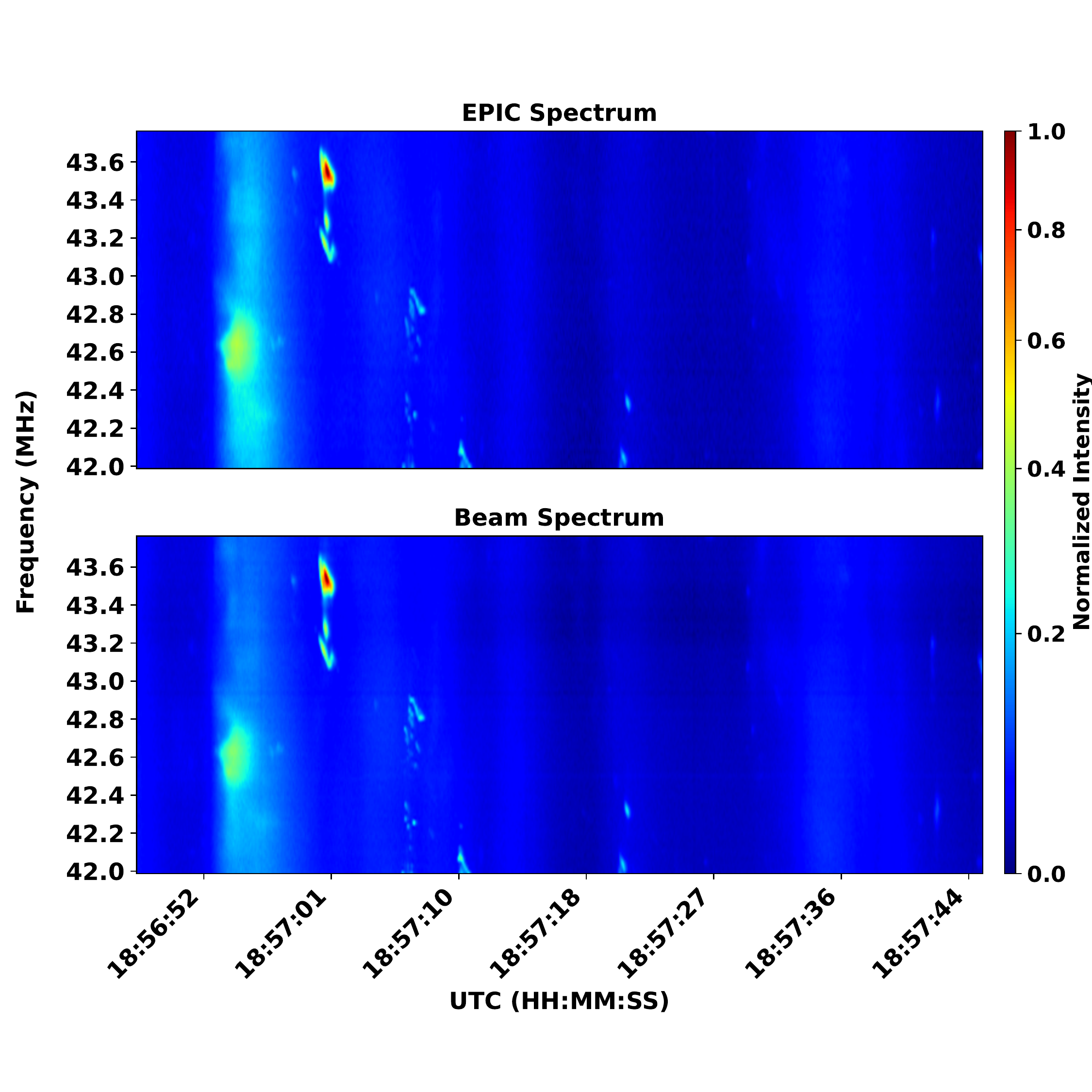}
    \caption{Comparison of dynamic spectra of the Sun from EPIC (top) with a simultaneous beam-formed measurement (bottom) using the standard LWA-SV pipeline. Both measurements span 1.8~MHz bandwidth and show the same emission features, which are indicative of faint Type III radio bursts and other fine structure. The EPIC data has a time resolution of 81.92~ms and frequency resolution of 25~kHz while the beam-formed data has a time resolution of $\sim$80~ms and frequency resolution $\sim$20~kHz.  The structures in both spectra agree well, although small difference are apparent (see Figure~\ref{fig:lcdiff}) that are explained by re-quantization in the standard LWA-SV pipeline that limited dynamic range and the slight differences in temporal and spectral resolution.}
    \label{fig:epicbeamsun}
\end{figure}

\begin{figure}
    \centering
	\includegraphics[width=\columnwidth]{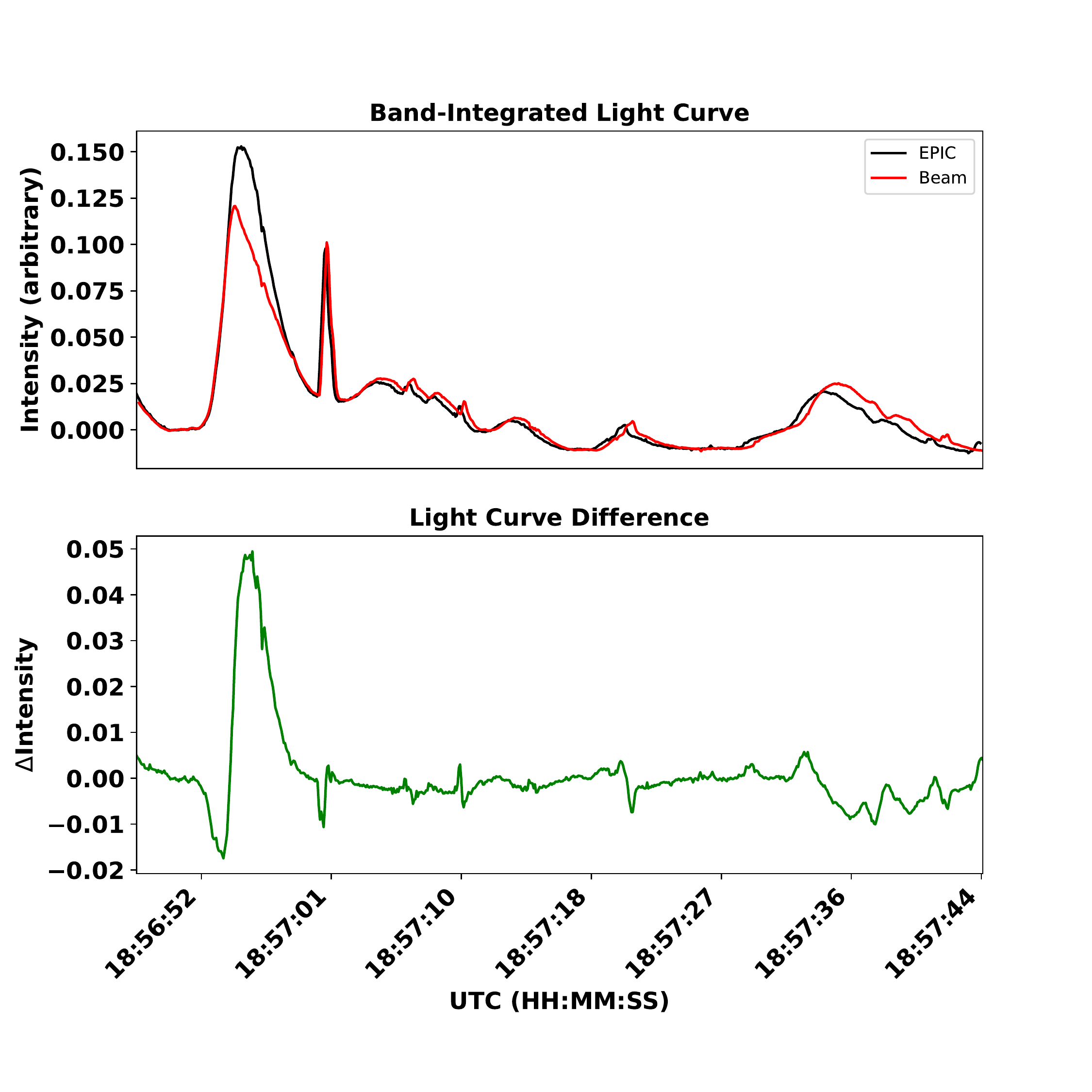}
    \caption{Comparison of band-integrated light curves corresponding to the dynamic spectra shown in Figure~\ref{fig:epicbeamsun}.  The top panel overlays the band-integrated light curves extracted from the EPIC images and from simultaneous beam-formed spectra created by the standard LWA-SV pipeline running commensally.  The bottom panel shows the difference in intensity between the two light curves. The total power is seen to track closely between the two cases.  The effects of the suppressed dynamic-range in the standard LWA-SV are visible as lower intensity in the beam-formed spectrum during the peak of the flare between 18:56:52 and 18:57:01.} 
    \label{fig:lcdiff}
\end{figure}

 As part of commissioning observations, we performed EPIC all-sky observing while simultaneously using the LWA-SV processing pipeline to acquire beam-formed observations on the Sun.  On 28~August 2021 between 18:00 and 21:00~UT, the Sun was in an active flaring phase with the beginning of the next solar maximum cycle.  Figure~\ref{fig:epicsun} displays a snapshot image of the sky from EPIC at the time of occurrence of the radio bursts. A comparison of the dynamic spectra from EPIC and the beam-formed observations is shown in Figure~\ref{fig:epicbeamsun} over a period of about one minute. The EPIC spectrum has been extracted from the images during the radio bursts from pixels corresponding to the location of the Sun. The dynamic spectra  display one-to-one correspondence of faint Type III radio bursts \citep{Nelson1985, Classen2002, Gopalswamy+2005} in addition to fine temporal and spectral structures. There are subtle differences between the two pipelines that are apparent in the band-integrated light curves shown in Figure~\ref{fig:lcdiff}. These are mainly due to the limited dynamic range of the beam-formed data from re-quantization in the standard LWA-SV pipeline and also the slightly different temporal and spectral resolutions between the two datasets.  Small differences in the effective sky region of the pixels extracted from EPIC images compared to the beam created by the standard LWA-SV processing pipeline also contribute to these effects. Overall, the observations of the Sun validate the EPIC processing and demonstrate its ability to discern sub-second duration transients and localize them to within the resolution limit of the telescope. We note  the standard LWA-SV pipeline is restricted on the number of beams that can be formed with the array. In contrast, EPIC yields dynamic spectra for all pixels in its all-sky images without compromising on frequency resolution, time resolution, or bandwidth.

\subsection{Bright low-DM Pulsars} 

\begin{figure}
    \centering
	\includegraphics[width=\columnwidth]{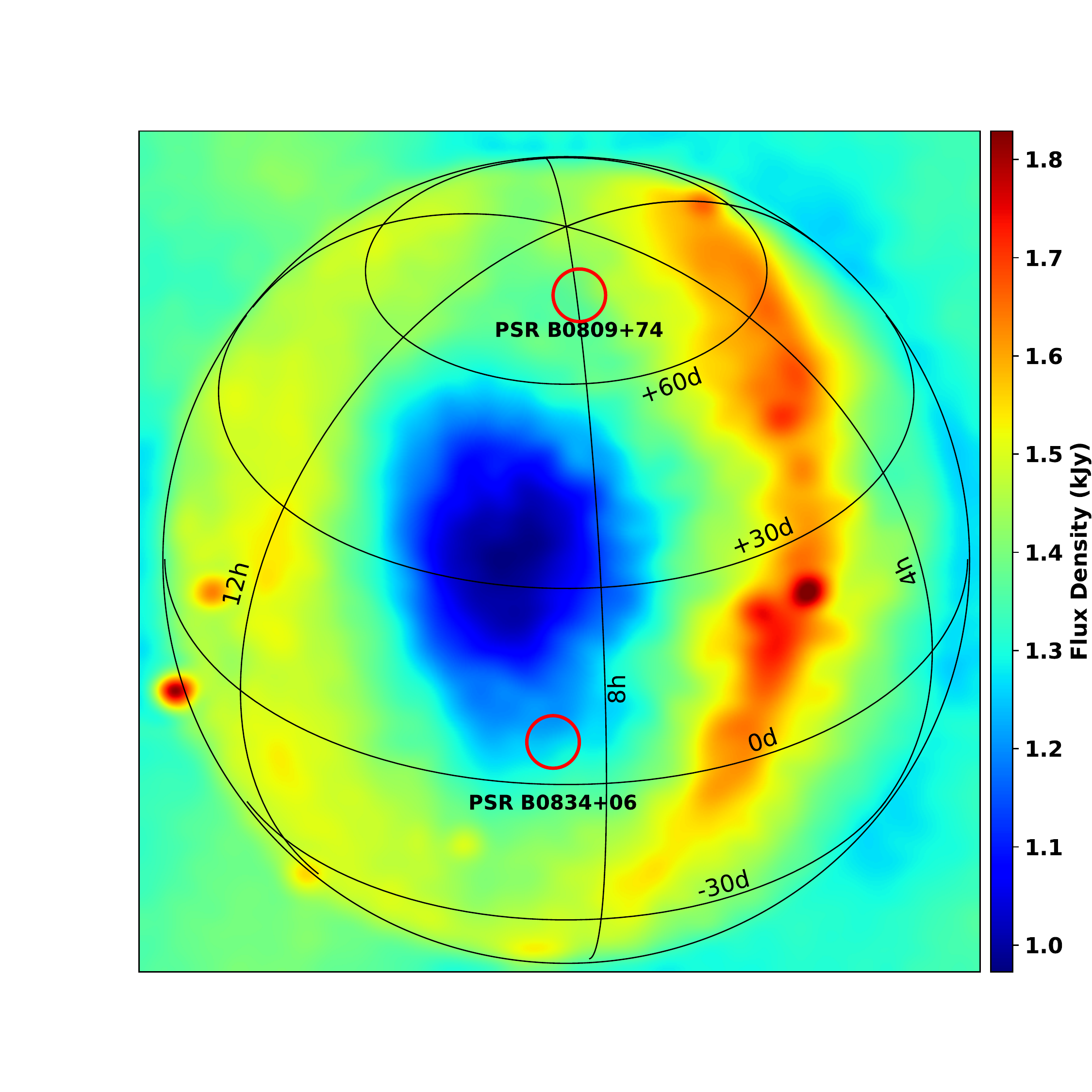}
	\includegraphics[width=\columnwidth]{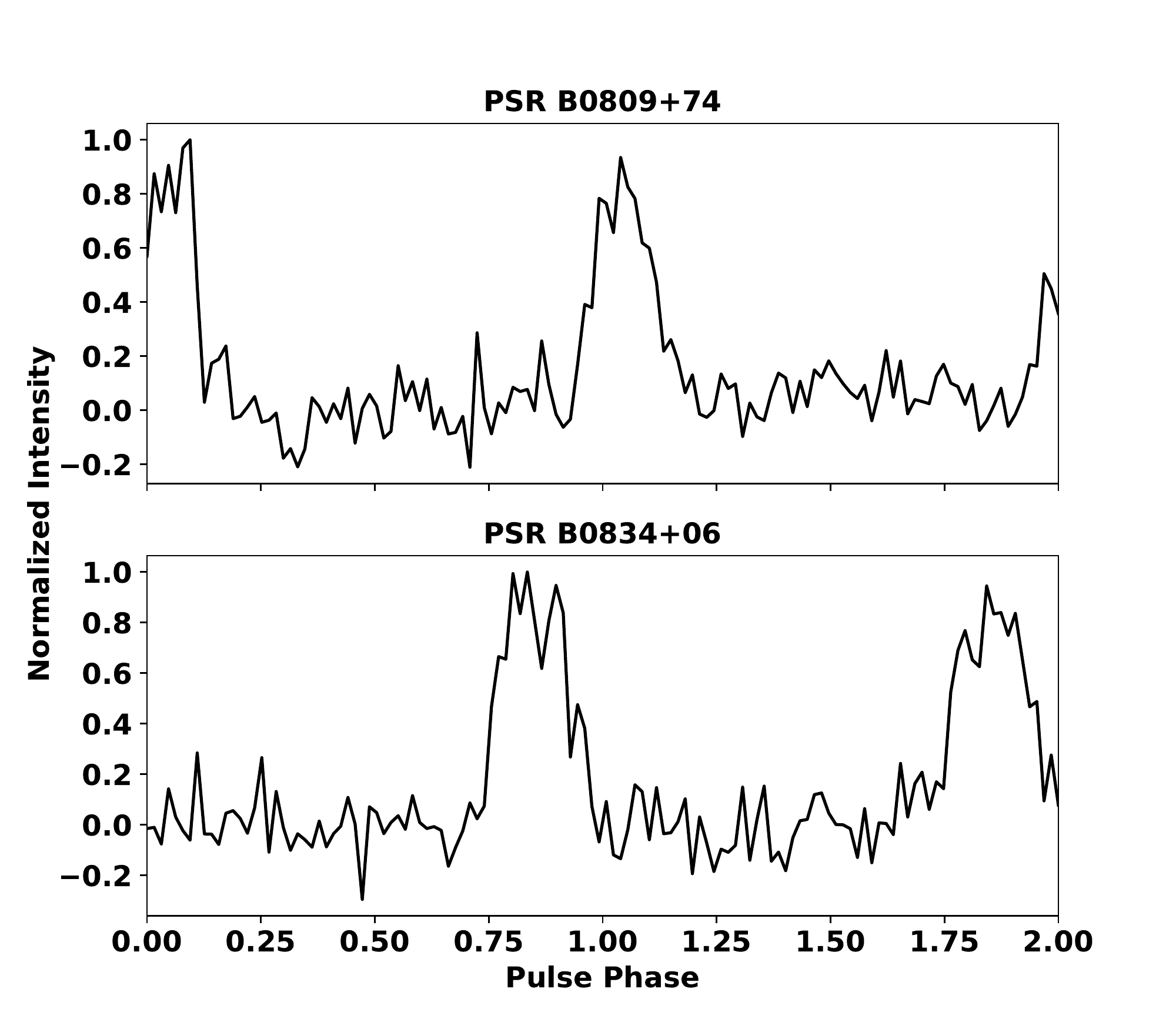}
    \caption{Snapshot full-sky image (top) from EPIC marking the simultaneous observation of the pulsars PSR~B0809+74 and PSR~B0834+06.  The integrated pulse profiles (bottom) derived from the light curves extracted from the EPIC images over a period of 45~minutes show the expected pulses. The observation used 1.8~MHz bandwidth centered at 43~MHz and the folded profile is binned to cover two pulse periods in 128~phase-bins.}
    \label{fig:prof}
\end{figure}

\begin{figure}
    \centering
	\includegraphics[height=0.35\textwidth, width=1.1\columnwidth, center]{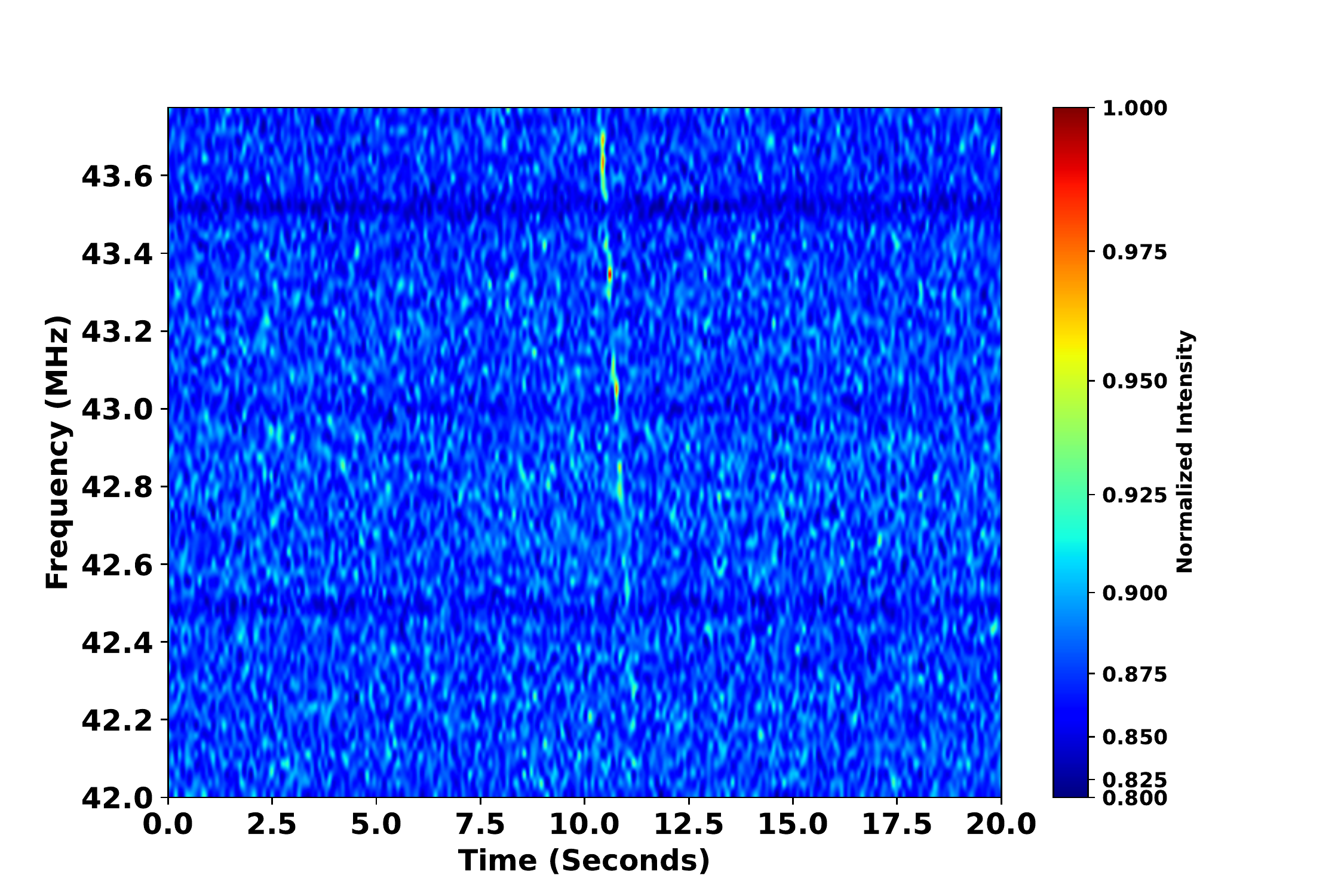}
    \caption{Dynamic spectrum extracted from EPIC imaging showing a giant pulse from the bright pulsar PSR~B1133+16 that drifts across the band due to dispersion.  The observations were acquired with frequency resolution of 25~kHz and a time resolution of 81.92~ms.}
    \label{fig:GP}
\end{figure}

We observed a few \enquote{bright} pulsars \citep{Bondonneau+2020}, listed in Table~\ref{Tab2:pulsarlist}, with low dispersion measures (DM$<20$~pc~cm$^{-3}$; \citealt{Davidson1969, Kulkarni2020}) close to their meridian transit at LWA-SV at 43~MHz. Each observing session lasted for about 45~minutes in order to observe many pulses to achieve reasonable signal-to-noise ratio (SNR) for the average profile \citep{Burns1969}. The light curves are extracted from the images for every frequency channel and folded corresponding to the period of the pulsar. The dispersive delay is corrected across the frequency band before integrating to produce the average pulse profile. The pulse profiles observed for PSR~B0834+06 and PSR~0809+74 are displayed in Figure~\ref{fig:prof}. In addition to the folded profiles we also detect giant single pulses for PSR~B1133+16 during our single pulse search in the extracted light curves. Giant pulses are single pulses that are abnormally bright with high SNR and thought to be due to various physical processes within the pulsar magnetosphere and the surface of the pulsar \citep{Petrova2004, Petrova2006, Kuzmin2007}. Figure~\ref{fig:GP} shows a bright single pulse from PSR~B1133+16 that displays dispersive delay from high to low frequency. These observations further demonstrate the applicability of EPIC to detect and to follow-up known transient sources.  We note that the simultaneous processing of the wide-field of view enabled us to observe multiple pulsars (Figure~\ref{fig:prof}) during a single observing session which is a particularly useful feature for blind surveys and also for simultaneous monitoring of known transient locations.

\begin{table}
\normalsize
	\centering
	\caption{Bright Low-DM Pulsars Observed with EPIC}
	\label{Tab2:pulsarlist}
	\begin{tabular}{lcc}
		\hline
		Name  & Dispersion Measure  & Period \\
		~ & [pc cm$^{-3}$] & [s] \\
		\hline
		PSR B0809+74 & 5.75066 & 1.292241446862 \\
		PSR B0834+06 &  12.8640 & 1.2737682915785 \\
		PSR B1133+16 & 4.84066 & 1.187913065936 \\
		\hline
	\end{tabular}
\end{table}

\subsection{Ionospheric Scintillation}

\begin{figure}
	\includegraphics[width=\columnwidth]{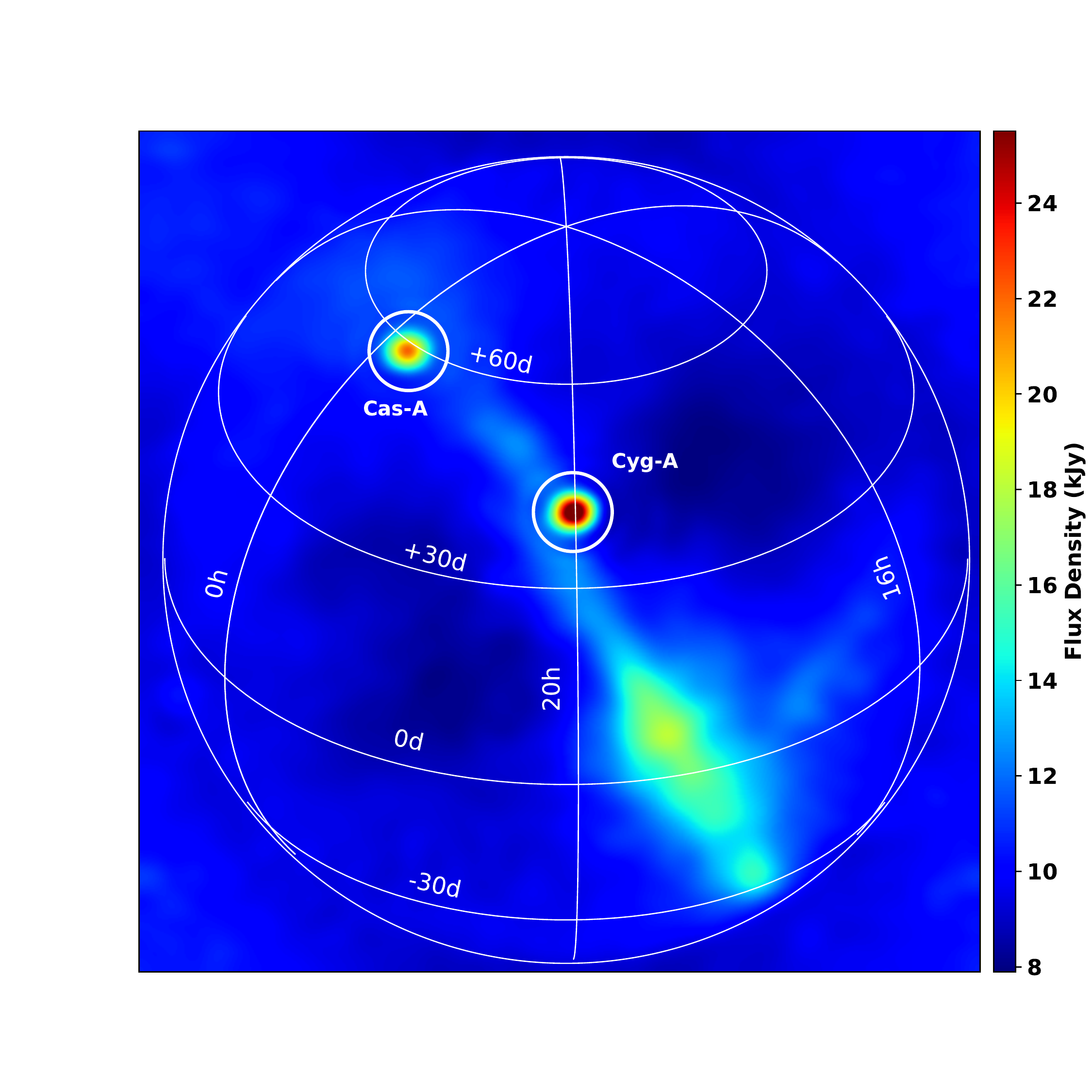}
    \includegraphics[width=1.2\columnwidth]{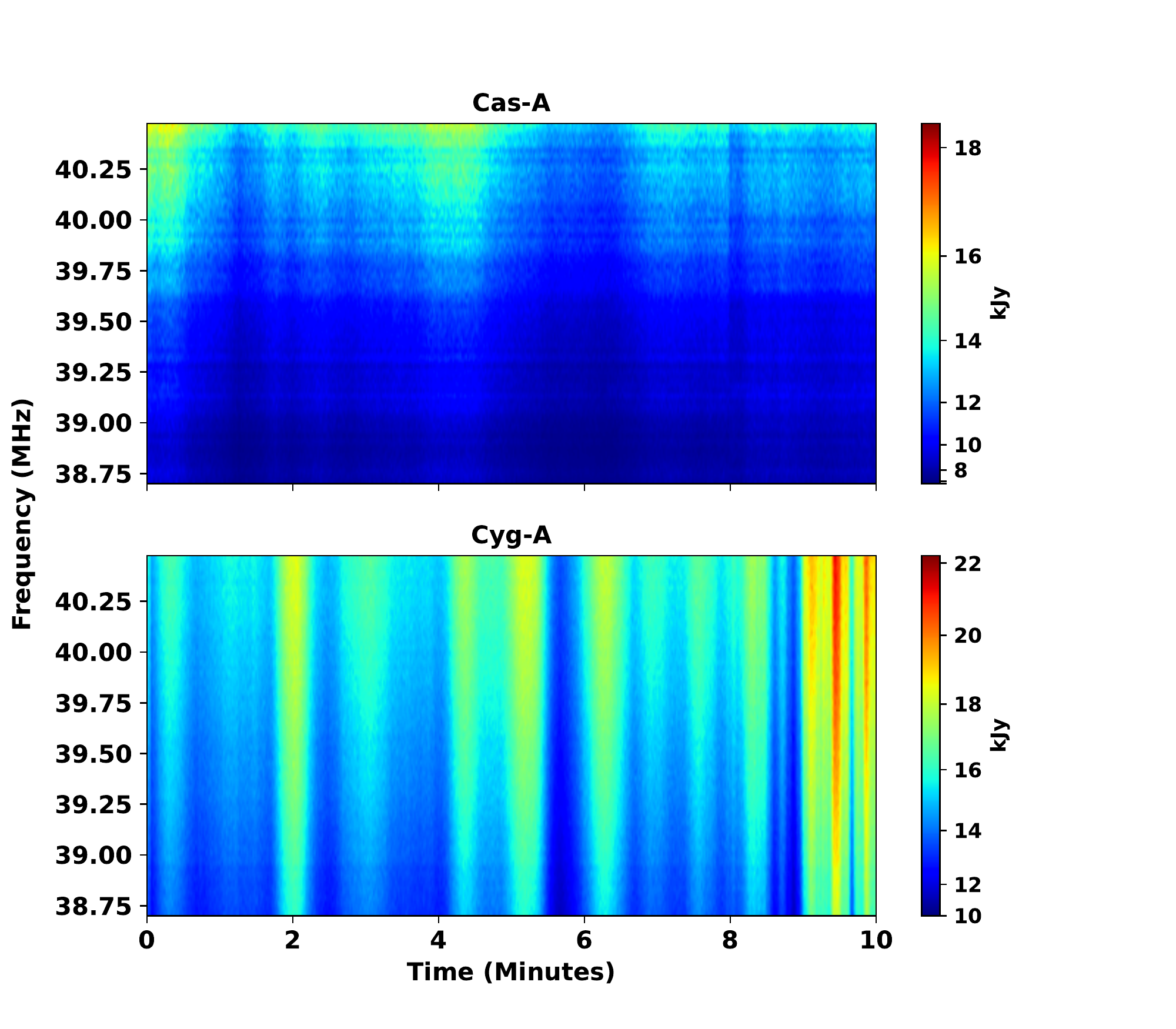}
    \caption{Snapshot full-sky image (top) from EPIC during the meridian transit of Cyg-A. The Galactic Centre is visible to the South and Cas-A to the North. Data were integrated for 820~ms over 250~kHz centred at 39.5~MHz.  Dynamic spectra (bottom panels) from EPIC for Cas-A and Cyg-A display ionospheric scintillation patterns.  The data have time resolution of 81.92~ms and frequency resolution of 25~kHz. Source intensities are estimated using the models for Cyg-A and Cas-A derived in \protect\cite{PB2017}.}
    \label{fig:ionsci}
\end{figure}

We also performed observations during the meridian transit of the bright radio source Cyg-A during the night of 20~September~2021. Figure~\ref{fig:ionsci} shows the full-sky image using EPIC. Dynamic spectra are extracted from pixels corresponding to the source locations of Cyg-A and Cas-A from a sequence over a 10~minute duration over a bandwidth of 1.8~MHz at a frequency resolution of 25~kHz. The dynamic spectra shown in the bottom panel of Figure~\ref{fig:ionsci} display distinct intensity fluctuation patterns on scales of a few seconds. This intensity variation is consistent with ionospheric scintillation caused by refractive scattering of radio waves as they propagate through the turbulent ionospheric plasma \citep{Crane1977, Yeh1982, Kintner+2007}. 

Observations of scintillating sources \citep{Obenberger+2015} to study the ionosphere and the interplanetary medium are important applications of EPIC. The wide-field of view and the high-cadence of EPIC is useful to extract dynamic spectra for a number of scintillating sources at various angular distances to the Sun to measure the \enquote{interplanetary scintillation} (IPS : \citealt{Hewish+1964, Coles1978}) in order to study and understand solar wind turbulence \citep{Ananth+1980, Fallows2013}. Also, the planned implementation of EPIC on other stations \citep{Taylor+2019} would transform the LWA as a multi-station IPS array which is useful to isolate ionospheric and interplanetary scintillation patterns as was demonstrated by \cite{Fallows+2016}. 

\section{Conclusion}

 Following the first GPU implementation and deployment of EPIC by \cite{Kent+2019}, we have performed code optimizations to specific 
 Bifrost blocks in EPIC and addressed some of the early computational challenges from the first deployment. Coupled with updated hardware, we have achieved an eighteen-fold increase in the bandwidth processed on a single GPU in real-time. EPIC is now commissioned as a commensal back-end at the LWA-SV station and runs on one server imaging a total bandwidth of $\sim$3.6~MHz (1.8~MHZ per GPU) configured with a default integration time of 81.92~ms. 

 We have demonstrated the effectiveness of EPIC through observations of solar radio bursts, long-period pulsars PSR~B0834+06 and PSR~0809+74, giant pulses from PSR~B1133+16, and ionospheric scintillation of the bright sources Cyg-A and Cas-A. These observations were complemented by simultaneious beam-formed observations from the digital system at LWA-SV for verification. The observation of weak solar bursts and pulsar giant pulses demonstrate the usefulness of EPIC as a potential transient imaging back-end for compact arrays.

 Future plans for EPIC include the addition of new compute nodes to the LWA-SV processing cluster with the aim to optimally cover a broader bandwidth of $\sim$20~MHz using a larger grid-size of 256$\times$256 able to match the angular resolution of the telescope at higher frequencies up to 80~MHz. We also plan to augment the imaging capabilities of EPIC by integrating additional modules that enable real-time calibration, RFI-detection and flagging, de-dispersion, and transient detection with a science focus to detect and localize impulsive transient sources like FRBs. Work is under progress to include the above features to make EPIC a generic science capable back-end that could be used by the community. As we upgrade EPIC, we will continue to operate in the commensal mode on the LWA-SV for all-sky imaging, performing blind-searches for low frequency transients and monitoring known sources within the sensitivity limits of the array.

\section*{Acknowledgements}

This work is supported by National Science Foundation awards AST-1710719 and AST-1711164 and by NASA Solar System Exploration Research Virtual Institute cooperative agreement number 80ARC017M0006. Construction of the LWA has been supported by the Office of Naval Research under Contract N00014-07-C-0147 and by the AFOSR. Support for operations and continuing development of the LWA is provided by the Air Force Research Laboratory and the National Science Foundation under grants AST-1835400 and AGS-1708855.
\section*{Data Availability}

The data underlying this article will be shared on reasonable request to the corresponding author.



\bibliographystyle{mnras}
\bibliography{reference} 








\bsp	
\label{lastpage}
\end{document}